\documentclass{article}
\usepackage{amsmath,amssymb}
\usepackage{graphicx}
\usepackage{hyperref}
\usepackage{cite}
\usepackage{multicol}
\usepackage{caption}
\usepackage{subcaption}
\usepackage{enumitem}
\usepackage[margin=1in]{geometry}
\usepackage{authblk}
\usepackage{float}

\title{Chainless Apps: A Modular Framework for Building Apps with Web2 Capability and Web3 Trust}
\author[1]{Brian Seong}
\author[1]{Paul Gebheim}
\affil[1]{Polygon Labs}
\date{}

\begin{document}

\maketitle

\begin{abstract}

Modern blockchain applications are often constrained by a trade-off between user experience and trust. Chainless Apps present a new paradigm of application architecture that separates execution, trust, bridging, and settlement into distinct compostable layers. This enables app-specific sequencing, verifiable off-chain computation, chain-agnostic asset and message routing via Agglayer, and finality on Ethereum — resulting in fast Web2-like UX with Web3-grade verifiability. Although consensus mechanisms have historically underpinned verifiable computation, the advent of zkVMs and decentralized validation services opens up new trust models for developers. Chainless Apps leverage this evolution to offer modular, scalable applications that maintain interoperability with the broader blockchain ecosystem while allowing domain-specific trade-offs.

\end{abstract}


\vspace{1em}

\begin{multicols}{2}
\section{Introduction}

Modern blockchain apps often face a trade-off between user experience and trust. Users must manage wallets, pay gas, and navigate bridges—barriers that degrade the seamlessness of Web3. Meanwhile, traditional blockchains require global re-execution of every transaction, ensuring strong trust but limiting scalability. Adding decentralization boosts security \& transparency, not throughput, since all nodes do the same work. This creates a UX gap that dApps struggle to overcome.

Verifiable compute addresses this by decoupling execution from verification. Off-chain computation is performed by a single prover or small group, who then supply a proof (e.g., ZK or TEE attestation) that others can cheaply verify. This approach preserves trustlessness while massively improving throughput.

Chainless Apps represent a new paradigm at the intersection of verifiable compute, modular infrastructure, and cross-chain interoperability. They are modular, chain-agnostic applications designed to deliver Web2-like performance and UX while preserving Web3-grade trustlessness. In a Chainless App, most application logic executes off-chain for speed and scalability in a VM-agnostic environment, then is verified and settled on-chain for security and settlement.

The settlement can be tied to a particular chain, but that tie-in brings the shortcomings of a particular chain (limited ecosystem, settlement times) to the fore.  Thus, these applications leverage shared cross-chain infrastructure, specifically Agglayer, for unified messaging, asset bridging, and settlement. Agglayer empowers apps to treat entire Web3 as a unified ecosystem (a la Web2) while abstracting away the complexity of multichain interaction, including gas fees and bridging logic, providing a seamless user experience across networks. 

Chainless Apps also follow an app-specific sequencing model\cite{Agarwal2022}, where each application behaves like its own sovereign execution chain, controlling its own transaction ordering and state evolution independently off-chain. This design enables faster execution, domain-optimized state machines, and trust-minimized verifiability without relying on global consensus for every transaction.

Chainless Apps unlock a new class of user-facing applications that were previously infeasible under traditional blockchain constraints — such as real-time trading engines, fast-paced multiplayer games, or socially interactive experiences that require responsiveness, privacy, and verifiability without incurring the latency or UX friction of on-chain interaction. This framework generalizes ideas previously explored in the [A-Z]app Paradigm\cite{Seong2024}. Recent systems like vApps~\cite{vApps2024} also explore modular application architectures that combine high-throughput execution with verifiable computation, reinforcing the broader trend toward app-specific design patterns.

Chainless Apps aim to feel like modern web apps—fast, intuitive, and accessible—while retaining the trust guarantees of decentralized systems. By combining off-chain execution with on-chain verification, and separating concerns across execution, trust, bridging, and settlement layers, developers can tailor trust models to their needs. To ground this architecture in a concrete example, we introduce zkSpot, a proof-of-concept high-performance spot trading application built using the Chainless App design. zkSpot offers centralized exchange-like speed with verifiable execution and seamless cross-chain asset support. The rest of this paper outlines the Chainless App architecture, explores a trading-focused implementation, and considers future implications for scalable decentralized applications.

\section{Background}

The emergence of modular blockchain design\cite{Kim2023Modularity} and verifiable compute has laid the foundation for a new class of applications that combine performance, security, and interoperability. This section introduces the two core concepts that underpin the Chainless App framework: verifiable computation as a scalable execution model, and Agglayer as the infrastructure enabling seamless cross-chain coordination.

\subsection{Verifiable Compute}
At the heart of Chainless Apps is the notion of verifiable compute. Verifiable compute refers to the ability for a system to prove that a given computation was executed correctly, such that others do not need to re-run that computation in full to trust the result. 

In practice, verifiable compute is achieved by generating an attestation or proof that others can independently verify. If valid, the result can be accepted without re-executing the computation. A common approach is using zero-knowledge proofs (ZKPs), where an off-chain executor runs application logic and produces a succinct proof (e.g., SNARK\cite{BenSasson2013} or STARK\cite{BenSasson2018}) attesting to its correctness. These proofs are cheap to verify and imply correct state transitions. This model powers zkRollups and zkVMs, enabling Ethereum to validate large batches of transactions with minimal on-chain cost.

Another method uses Trusted Execution Environments (TEEs), secure hardware zones (e.g., Intel SGX, ARM TrustZone) that ensure computation integrity and generate attestations. Though not trustless, since it relys on hardware and manufacturers, TEEs offer high performance and a significantly higher bar for cheating. TEEs can also emit execution traces that are later verified by zk-proofs for stronger guarantees.

Verifiable compute can also rely on consensus: small validator committees re-execute logic and agree on the outcome. This model, used in optimistic rollups\cite{Kalodner2018} and some bridges, assumes at least one honest participant to detect fraud. While not as strong as ZKPs, consensus-based verification remains robust when backed by slashing and incentives. 

Chainless Apps are designed to be flexible in what trust model they use. The developer can choose a mechanism (or even combine multiple) that best fits their application’s requirements for security, cost, and performance. In all cases, the goal is the same: decouple execution from on-chain consensus by doing work off-chain and then proving to the chain that the work was done correctly. This lays the foundation for massive scalability gains while retaining the security of the underlying blockchain.

\subsection{Agglayer - Interoperability \& Settlement Layer}

The future of blockchain scalability is modular and multichain. Rather than building solely on a Layer-1, developers are adopting Layer-2s\cite{Buterin2021Rollups}, app-specific chains, and modular blockchains, which decouples execution, data availability\cite{AlBassam2019}, and settlements. Base layers like Ethereum provide security, while execution is offloaded to faster secondary environments.

A key enabler of this vision is chain abstraction where users shouldn’t need to manage bridges or switch networks. Agglayer provides the infrastructure for this abstraction. It unifies liquidity, users, and state across chains by treating multiple networks as a single, coherent system. At the heart of Agglayer is the Unified Bridge, a standardized interface and communication protocol deployed on all Agglayer connected chains. It enables seamless asset and message routing across chains without wrapping or custom bridges, simplifying both integration and liquidity flows. 

To maintain cross-chain safety, Agglayer enforces pessimistic proofs, an invariant check that prevent over-withdrawals. This mechanism acts as a built-in safeguard against exploits, ensuring that no chain can extract more than it deposited. Agglayer also enables proof aggregation. Instead of each chain posting its validity proof to Ethereum individually, Agglayer bundles multiple proofs into one aggregate submission. This reduces settlement costs, improves efficiency, and synchronizes finality across chains.

For real-time use cases, Agglayer will supports fast interop, sub-second messaging across chains, as well as third-party shared sequencing, which enables cross-domain coordination\cite{Han2024SharedSequencer}. These features allow apps to trigger logic or finalize state on other networks almost instantly, unlocking responsive, multi-chain use cases such as trading, gaming, and composable DeFi.

With Agglayer providing the infrastructure for fast, secure, and seamless multichain coordination, Chainless Apps are uniquely positioned to thrive. They combine off-chain execution, modular trust, and unified cross-chain settlement into a cohesive framework. By abstracting away the complexities of chain management and verification, developers can focus on application logic and user experience, while end users enjoy fast, gasless, and intuitive interactions. In the following sections, we introduce the Chainless App architecture in detail, covering its layered design, trust model flexibility, and real-world implementation examples.

\end{multicols}

\begin{figure*}[ht]
    \centering
    \includegraphics[width=0.65\textwidth]{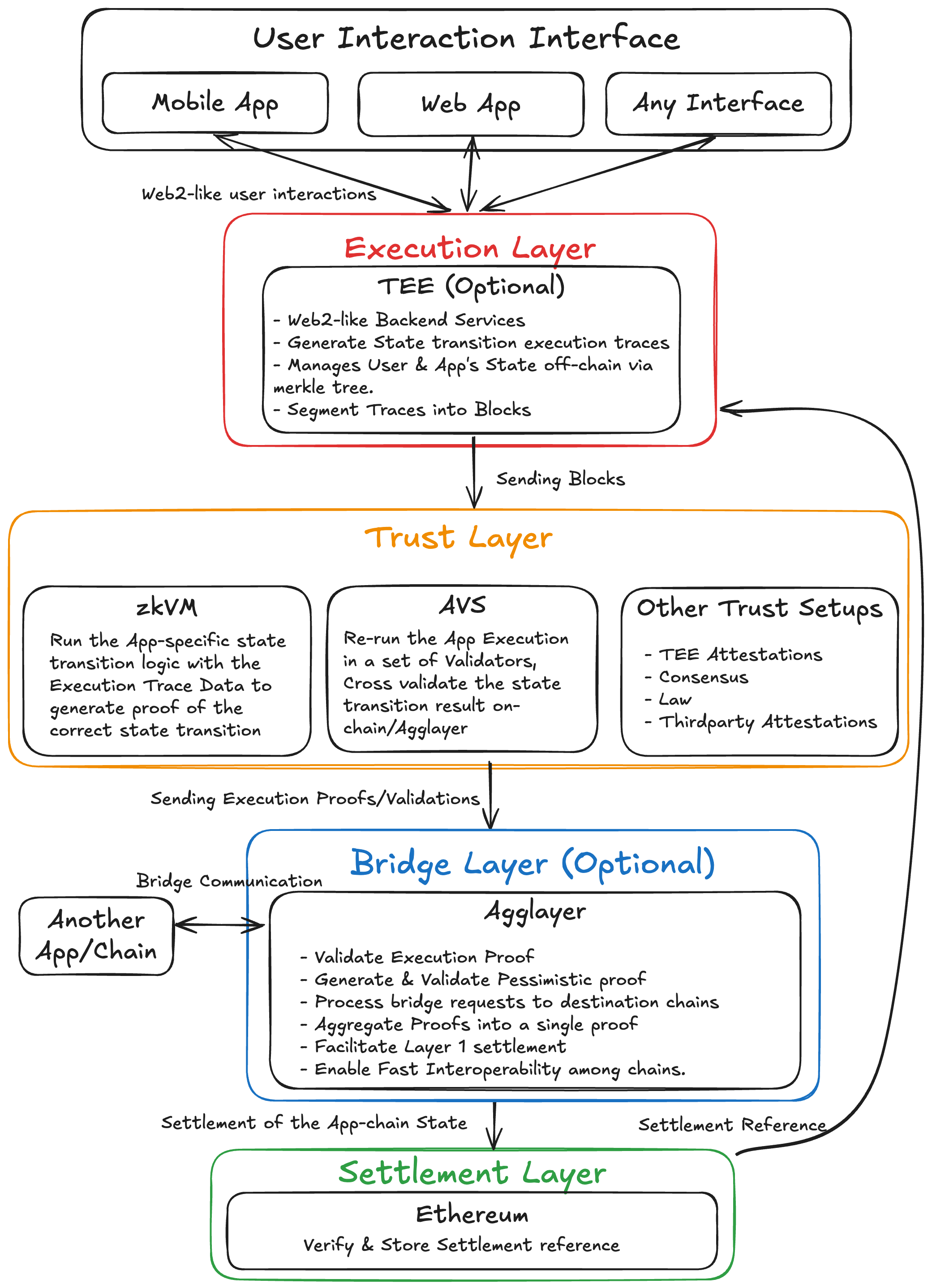}
    \caption{High-level architecture of a Chainless App. It shows user interaction through Web2-like interfaces, off-chain execution in TEEs, validation via trust layers (ZKVMs, AVSs), and cross-chain bridging and settlement via the Agglayer.}
    \label{fig:chainless}
\end{figure*}

\begin{multicols}{2}

\section{Chainless App Architecture}
A Chainless App is organized into multiple layers, each responsible for a different aspect of the application’s operation. This layered design (illustrated in \ref{fig:chainless}) decouples the fast but off-chain execution from the trust enforcement and blockchain interaction. Layers include: (i) an Execution Layer where the application’s business logic runs and the state is managed off-chain, (ii) a Trust Layer which verifies the accuracy of the execution using one or more methods of verifiable computation, (iii) an Interoperability Layer (built on the Agglayer) which handles cross-chain communication and asset movement, and (iv) a Settlement Layer that finalizes the state of the application on a base blockchain for long-term security and dispute resolution. By separating these concerns, Chainless Apps achieve modularity and flexibility – each layer can potentially be implemented or configured in different ways without affecting the others. Below, we describe each layer in detail and how they interoperate to provide a seamless yet trustless experience.

\subsection{Execution Layer}

The Execution Layer of a Chainless App is responsible for running the application's core logic and maintaining its off-chain state, analogous to the backend of a traditional web app. Unlike traditional blockchains, Chainless Apps implement app-specific sequencing: each application defines its own state machine, handles transaction ordering independently, and logs state transitions off-chain. This decoupled model enables higher throughput and responsiveness than global blockchain consensus, while still enabling later verification and settlement.

By moving execution off-chain, apps can process many operations per second without being limited by block times or gas costs. The Execution Layer runs in a controlled environment—such as a standard runtime, WASM VM, or secure enclave—rather than as on-chain smart contracts. However, because this execution is untrusted by default, it must be paired with robust verification by the Trust Layer.

\textbf{Execution Trace and Application Logic}: The Execution Layer processes a stream of user inputs and records an ordered sequence of transitions, forming an execution trace. This trace represents how the application moved from one state to the next over each block or epoch. While app logic can be implemented in any language or environment, it can be abstractly modeled\cite{Kautuk2024Micro} as a Finite State Machine (FSM), where each transition is a deterministic function of the current state and an input. This trace becomes the canonical record of computation and is committed cryptographically (e.g., via Merkle roots or hash chains).

The Trust Layer later consumes this trace for verification. zkVMs can deterministically replay the computation and generate succinct proofs of correctness. Similarly, AVS or committee-based systems can re-execute the logic to verify consistency. This structure ensures computation is auditable, deterministic, and anchored in verifiable commitments.

\textbf{TEE Execution}: Optionally, Chainless Apps may run execution inside a Trusted Execution Environment (TEE) to ensure low-latency, high-throughput operation with integrity guarantees. A TEE (e.g., Intel SGX or AMD SEV) enables application logic to run at near-native speed while protecting internal state from tampering—even by privileged operators. This model provides CEX-like responsiveness, allowing a trading app to match orders and update balances in milliseconds. TEEs also mitigate MEV (Maximal Extractable Value) by sealing execution order against manipulation. As highlighted by Flashbots’ Rollup Boost framework\cite{Flashbots2024}, TEEs can preserve fairness in off-chain systems.

\textbf{Data Availability of Execution Trace}: Once generated, execution traces must be available for verification. Depending on the application’s privacy and cost needs, traces can be stored privately (with a hash committed on-chain), or publicly via decentralized data availability (DA) networks like Avail\cite{Avail2024}. This flexibility allows developers to trade off between auditability and overhead.

\textbf{Execution Security}: While TEEs offer performance and integrity, they are not foolproof. A compromised enclave—through side-channel attacks, bugs, or insider access—can produce invalid results with valid-looking attestations. Therefore, Chainless Apps treat the Execution Layer as untrusted until verified. Every output must be validated by the Trust Layer using the execution trace.

To optimize batching and checkpointing, execution traces are often segmented into blocks or epochs, each committed with a Merkle root or hash. Periodic checkpoints can be generated every $\mu$ transactions or $\nu$ seconds, aligned with proof generation and bridge sync intervals.

This design allows Chainless Apps to execute thousands of transactions off-chain while preserving verifiability, correctness, and finality when anchored on-chain through Agglayer.

In essence, the Execution Layer behaves like a rollup sequencer—but is application-specific and can configured to run in a secure enclave. It accumulates a sequence of transactions and produces verifiable evidence of correct execution, ready for downstream validation and settlement.

\subsection{Trust Layer}
The Trust Layer is responsible for verifying that the Execution Layer’s outputs are correct and enforceable according to the application's rules. In other words, it provides the trustlessness guarantee: ensuring off-chain computation was performed faithfully before accepting state updates on-chain. Chainless Apps are modular at this layer, developers can select the verification method that best balances their application's security and performance needs.

\subsubsection{Trust Mechanisms}

Common trust mechanisms include zero-knowledge proof verification, autonomous verifiable service (AVS), replicated consensus, optimistic verification, or external trust assumptions. We focus first on the strongest model: zkVM-based verifiable compute.

\textbf{zkVM Verifiable Compute}: One of the most secure approaches is to use a zkVM (zero-knowledge Virtual Machine) such as Plonky3 based zkVMs\cite{Plonky32023} or RiscZero's zkVM\cite{RISCzero2023} to re-execute the application’s execution trace and produce a succinct proof of correctness. In this model, the off-chain execution trace is fed into the zkVM, which deterministically simulates the same transitions based on the application's logic. If execution was correct, the zkVM outputs a cryptographic proof (e.g., a SNARK or STARK) attesting that the final state was derived correctly.

This proof is small (hundreds of bytes) and extremely cheap to verify on-chain, while providing strong cryptographic assurances: if the proof verifies, the chance that an invalid computation passed is negligible under standard assumptions. Thus, zkVMs turn Chainless Apps into systems similar to validity rollups, where every off-chain computation is independently verifiable.

By integrating a zkVM, a Chainless App achieves maximum trustlessness: even if the Execution Layer (e.g., the TEE) were compromised, invalid state transitions would be detected and rejected by on-chain verification. The trade-off is that generating zero-knowledge proofs, especially for large or complex state transitions, can be resource-intensive. Luckily, the proving systems are rapidly becoming more efficient through advances in zk circuit design and hardware acceleration.

\textbf{Committee-Based Verifiable Compute}: Chainless Apps can leverage off-chain validator committees to validate execution traces and ensure correctness. In this model, a set of independent nodes either re-execute transactions (as seen in EigenLayer AVS\cite{EigenLayer2023} and Truebit\cite{Truebit2019}) or reach consensus on the resulting state (as in Tendermint or HotStuff).

Committee-based verification offers several benefits. It enables flexible, programmable validation of arbitrary application logic without requiring expensive zk-circuit design upfront. It also provides faster confirmation times compared to full zk-proving, making it attractive for applications where low latency is important. However, its security model varies based on configuration, some rely on slashing to deter dishonesty, others depend on the assumption of at least one honest validator willing to challenge fraud.

It is well-suited for apps seeking faster time-to-finality with decentralized security guarantees. One critical consideration remains data availability: validators must be able to access execution traces or input data efficiently, whether via public DA layers, private storage, or inline publication. The final choice between staking-based committees, consensus-based finality, or challenge-based fraud detection depends on the specific risk tolerance, cost sensitivity, and responsiveness required by the application.

\textbf{Operator Trust Compute}: At the most trust-dependent end of the spectrum, a Chainless App could operate without cryptographic or game-theoretic verification by simply asking users to trust the operator's off-chain results. This model prioritizes simplicity and speed, sometimes supplemented with legal contracts or insurance. While unsuitable for serious decentralized public-facing apps, Operator Trust can be acceptable for low-stakes use cases, private betas, early deployments, or institutional developers who has other web2 constraints for trust. Critically, even Operator Trust apps can leverage Agglayer for asset bridging and settlement, maintaining interoperability even if correctness relies purely on operator reputation. As applications mature, they can transition from Operator Trust to stronger verification models as needed.

\subsubsection{Trust Layer Design: Flexible Verification Models}

Chainless Apps allow developers to configure hybrid trust models depending on their security, performance, and cost requirements. Several verification strategies are possible:

\begin{itemize}[leftmargin=*, itemsep=0.5em]
    \item \textbf{Full Re-execution by Trust Layer}:  
    The Trust Layer deterministically replays the entire execution trace, validating every transaction step without trusting the TEE beyond transaction ordering.

    \item \textbf{TEE Sequencing + Partial Verification \& State Verification}:  
    First validate the TEE's attestation, then selectively re-verify critical transitions through zk-proofs or validator committees to balance trust and computational efficiency.

    \item \textbf{Sequencing + State Verification by Trust Layer}:  
    The Execution Layer processes transactions like a traditional Web2 server, but the resulting state transitions are independently verified by a Trust Layer component (e.g., zkVM, AVS), avoiding full re-execution to optimize performance.

    \item \textbf{No State Verification (Operator Trust Mode)}:  
    Forgo formal validation entirely and rely on operator trust, legal agreements, or insurance. Suitable only for low-risk, internal, early-stage deployments, or institutional usecases.

    \item \textbf{Other Custom Security Models}:  
    Applications are free to design their own hybrid or bespoke security models that match their specific threat models and operational needs.
\end{itemize}

\begin{figure}[H]
    \centering
    \includegraphics[width=0.49\textwidth]{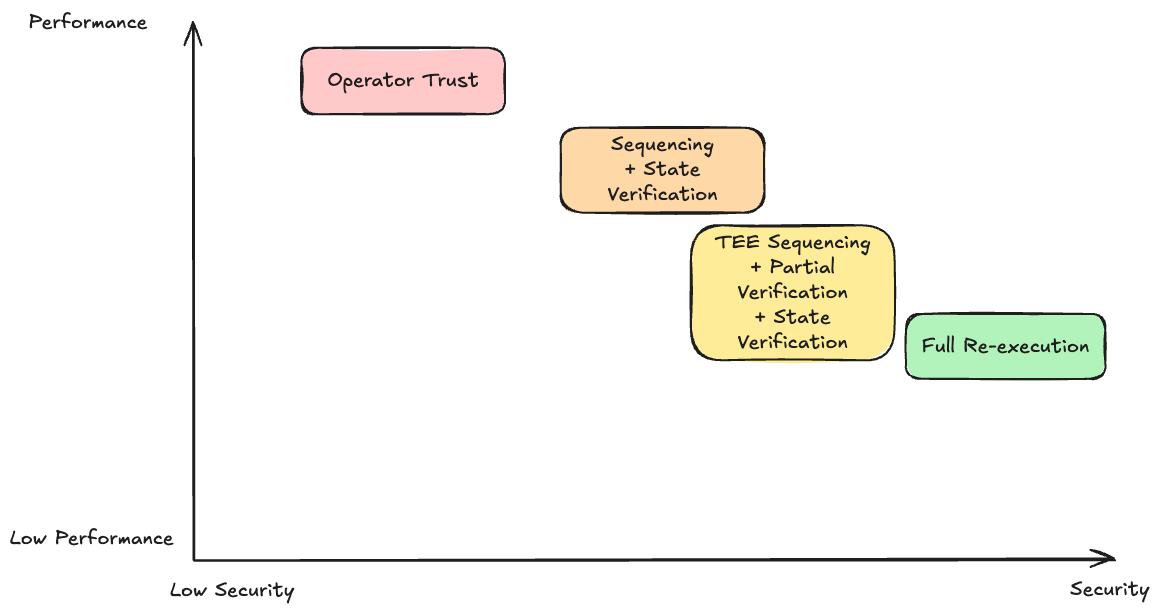}
    \caption{The more re-execution you run in the trust layer, the more secure it is, but the higher the performance overhead.}
    \label{fig:perf_vs_sec}
\end{figure}

The choice of trust model depends on the application's risk tolerance, performance targets, and maturity stage. High-value apps such as trading platforms or custody solutions may favor full zk-proof based verification, while latency-sensitive applications like gaming may initially leverage TEE-backed sequencing with periodic zk validation.

While this flexibility enables innovation, it also introduces a greater burden on users and auditors to evaluate the security posture of each Chainless App. Unlike traditional smart contract-based dApps or rollup SDKs where trust assumptions are more uniform, Chainless Apps require careful assessment of the chosen trust model, sequencing logic, and verification pathways. Future tooling for standardizing attestations, proof disclosures, or modular audits may help reduce this burden as the ecosystem matures.

Importantly, Chainless Apps can evolve their trust architecture over time, mixing, matching, or upgrading verification modules as technologies and threat models advance, without re-architect the core system. 

This flexibility is made possible because Agglayer isolates and secures interoperability independently of the internal trust assumptions of each app or chain. Through pessimistic proofs, Agglayer ensures that assets bridging out from a chain or Chainless App cannot be double spent or overdrawn, even if the internal security model of the application varies. In this model, each Chainless App is responsible for maintaining the authenticity of its internal state, while Agglayer guarantees the correctness and safety of cross-chain asset transfers and interactions. This separation of concerns allows diverse trust models to coexist while maintaining global interoperability guarantees across the ecosystem.

\subsection{Interoperability Layer Using Agglayer}

A critical feature of Chainless Apps is native interoperability—the ability to seamlessly move assets and messages across multiple blockchains without custom bridges or fragmented liquidity. Without a unified interoperability layer, every app would have to independently manage bridging logic, trust different networks, and risk security breaches. Chainless Apps solve this by integrating a dedicated Interoperability Layer built on Agglayer\cite{Polygon2024}.

Agglayer is an interoperability and settlement protocol that unifies liquidity, state, and messaging across diverse blockchain networks, while also facilitating settlement for Agglayer-connected chains. It abstracts away the fragmentation of cross-chain interaction, allowing applications to operate across ecosystems as if they were part of a single network. For Chainless Apps to operate on Agglayer, there are three main components to integrate with. First, the Unified Bridge provides a standardized asset and message bridging interface deployed across all connected chains, eliminating the need for chain-specific wrapping or custom bridges. Second, Pessimistic Proofs introduce a real-time token balance check that ensures no chain can withdraw more assets than it originally deposited, preventing common bridge exploits and maintaining liquidity safety across the system. Finally, AggKit is a lightweight developer toolkit that exposes core Agglayer functionality, enabling applications to easily integrate asset bridging and cross-chain messaging with minimal development overhead.

While Chainless Apps can technically operate without Agglayer, doing so reintroduces fragmentation in liquidity, messaging, and settlement logic. Agglayer provides an unified interop and settlement layer that simplifies cross-chain UX and protects asset invariants — enabling Chainless Apps to scale as first-class citizens of a multichain world.

\begin{figure}[H]
    \centering
    \includegraphics[width=0.45\textwidth]{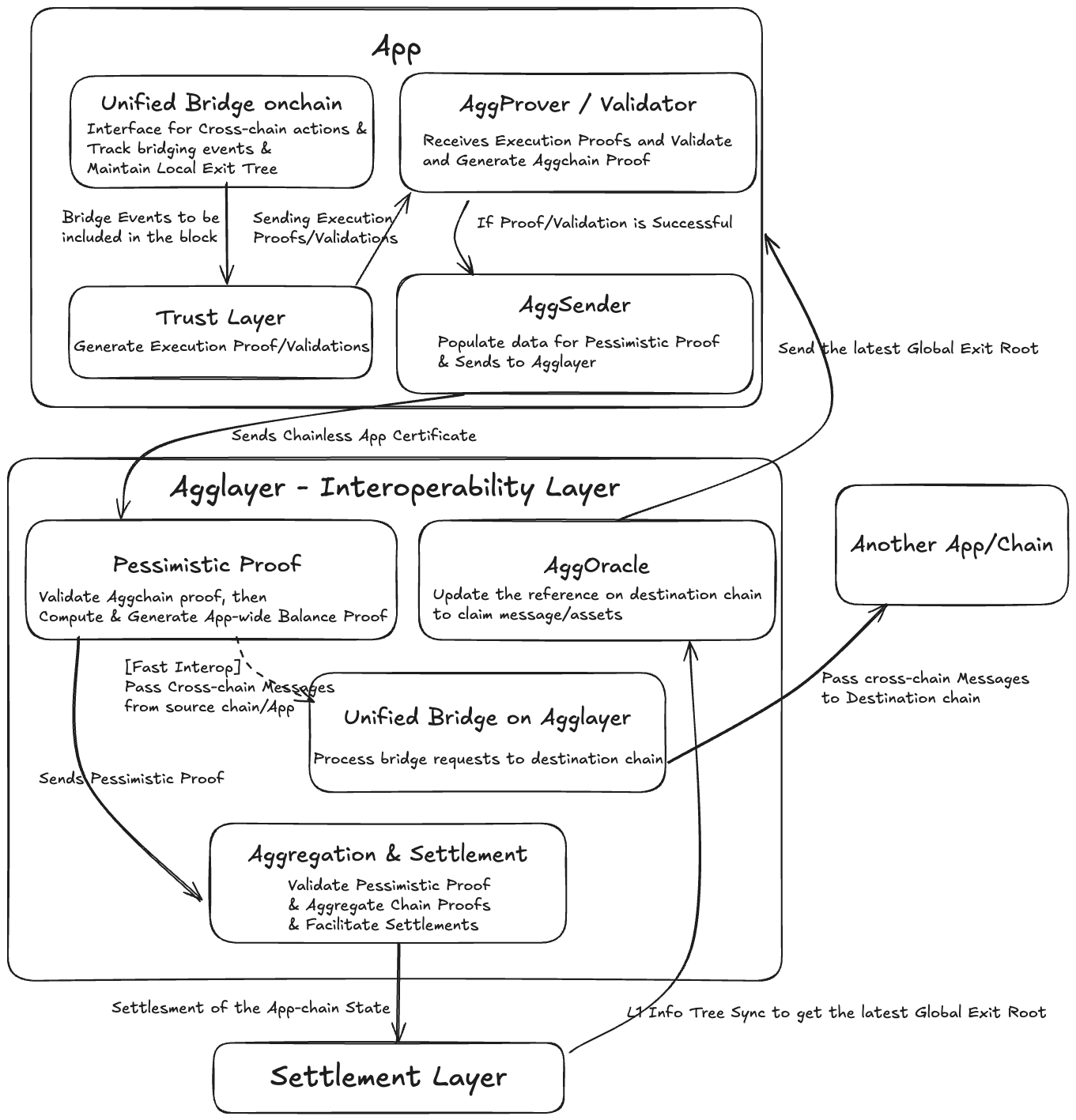}
    \caption{A detailed looks of Chainless-App's Agglayer integration.}
    \label{fig:Agglayer}
\end{figure}

By leveraging Agglayer, Chainless Apps gain access to secure and standardized cross-chain functionality. Apps can enable seamless deposits and withdrawals across chains, maintain unified liquidity tracking, and perform atomic cross-chain actions such as bridging assets and triggering smart contracts in a single user operation. Additionally, Agglayer’s Fast Interop feature enables sub-second cross-chain messaging, allowing latency-sensitive use cases like high-frequency trading, gaming, and composable DeFi to function smoothly across multiple chains.

Crucially, Agglayer separates internal application security from cross-chain settlement guarantees. Even if a Chainless App adopts a lighter internal trust model, such as relying primarily on TEE-backed execution, cross-chain asset movement remains protected by the pessimistic proof mechanism at the Agglayer level. This design ensures that regardless of an individual app’s internal verification strategy, the integrity of cross-chain transactions and liquidity remains safeguarded.

\begin{figure}[H]
    \centering
    \includegraphics[width=0.45\textwidth]{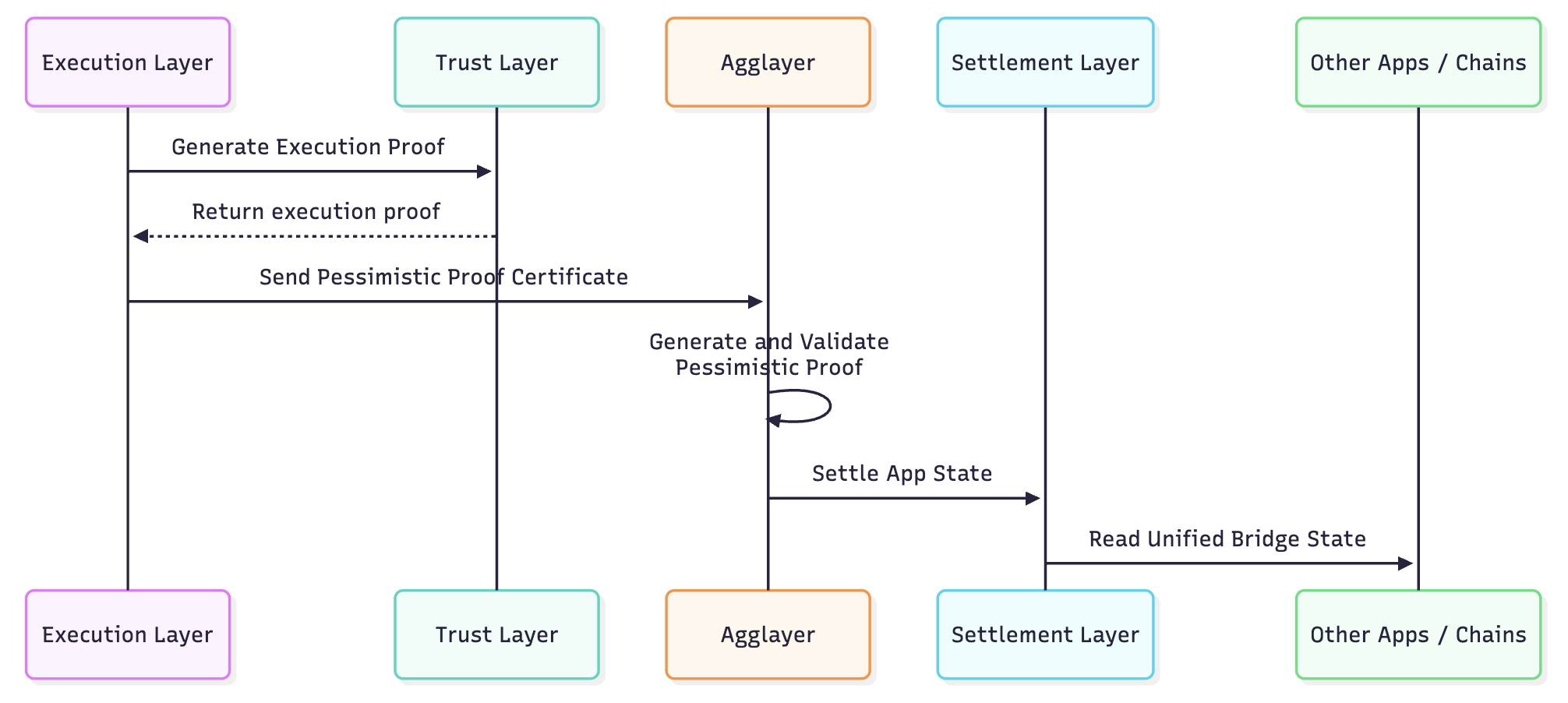}
    \caption{A highlevel sequence diagram of a Chainless App interacting with Agglayer}
    \label{fig:Agglayer_sequence}
\end{figure}

\subsection{Settlement Layer}

The Settlement Layer is the final step of the Chainless App architecture, anchoring off-chain execution and trust proofs onto a secure base blockchain such as Ethereum. While the Execution Layer runs transactions off-chain and the Trust Layer verifies their correctness, the Settlement Layer records the final application state immutably on-chain, ensuring full finality, dispute resolution, and long-term security.

In the Chainless model, settlement is typically handled by Agglayer’s smart contracts. When the Trust Layer generates a proof—such as a zk proof, validator signatures, or an optimistic validation, then an Agglayer node submits this data to the Settlement Layer. Along with state roots, bridge events are also recorded to ensure liquidity invariants are preserved across chains.

The Settlement Layer acts as the ultimate source of truth: once a state update is finalized, users can safely bridge out assets, reference balances, or resolve disputes based on the canonical on-chain state.

Although this paper assumes Ethereum as the default settlement layer, Chainless Apps can target any sufficiently secure base, such as Agglayer in the long run. In all cases, settlement anchors the app’s correctness and ensures that the app’s outcomes, balances, and interactions remain verifiable and trustless across Web3 ecosystems.
\section{What Chainless Apps Unlock}

Chainless Apps enable new classes of applications by decoupling execution from trust and settlement. With flexible execution environments and pluggable verification mechanisms, developers are no longer limited by the constraints of on-chain logic or rigid rollup architectures. This section showcases several application designs—ranging from high-performance trading platforms to verifiable games and decentralized social systems—that demonstrate the diversity of execution and trust configurations made possible by the Chainless App framework.

\subsection{zkSpot}

To illustrate the Chainless App architecture in practice, consider a high-performance spot trading application, which we will refer to as zkSpot\cite{zkHyperliquid2024} throughout this context. zkSpot aims to deliver the speed and user experience of a centralized exchange (CEX) — fast order matching, real-time updates, no visible gas fees — while retaining the trustlessness and self-custody of a decentralized exchange (DEX). In short: “Use it like a CEX, trust it like a DEX.”

\begin{figure}[H]
    \centering
    \includegraphics[width=0.4\textwidth]{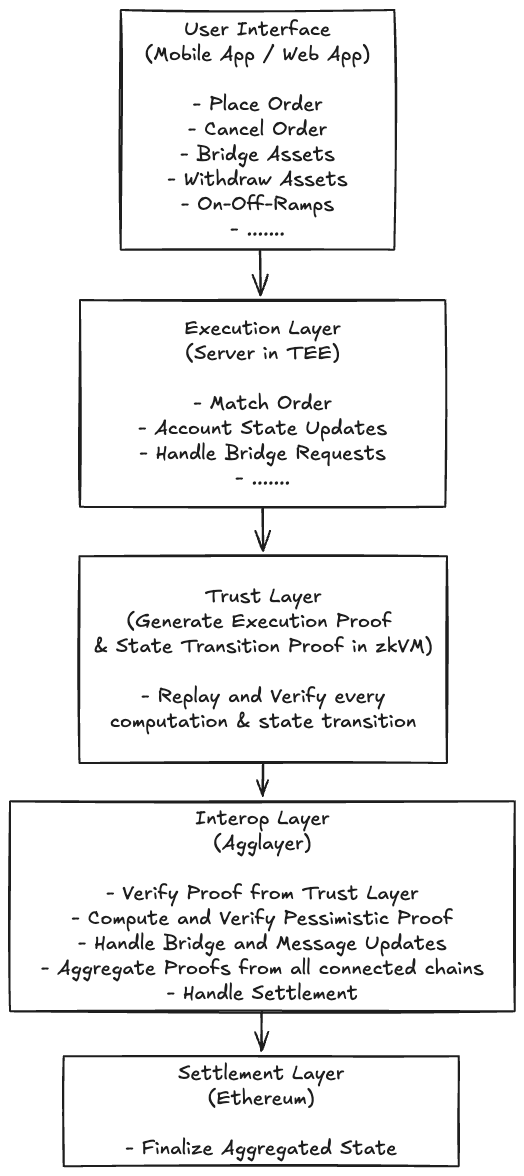}
    \caption{Flow of zkSpot execution, trust verification, cross-chain bridging, and settlement.}
    \label{fig:zkspot_flow}
\end{figure}

At \textbf{User Interface Layer}, traders interact through a Web2-style web application. They can place orders, deposit assets, and manage balances without signing blockchain transactions in real-time. Onboarding is simplified through optional embedded wallets, with Agglayer handling the underlying cross-chain deposits and withdrawals.

\textbf{The Execution Layer} runs the trading engine inside a Trusted Execution Environment (TEE), maintaining an off-chain central limit order book (CLOB). Trades are matched and balances updated within milliseconds inside the enclave, with execution logs generated for each batch. These logs are cryptographically attested to ensure execution integrity and enforce transaction ordering guarantees.

\textbf{The Trust Layer} verifies each batch using a zkVM (such as Succinct's SP1\cite{SuccinctLabs2023SP1}). The zkVM deterministically replays the trade logic based on the execution trace and generates a succinct zk-proof attesting that all matching and balance updates were performed correctly. This proof is then submitted to Agglayer for confirmation.

At \textbf{the Interoperability Layer}, Agglayer validates the zk-proof and updates the unified cross-chain state. Verified state transitions allow users to bridge their updated balances to any Agglayer-connected chain.

Finally, \textbf{the Settlement Layer} records zkSpot’s finalized state on Ethereum. The Agglayer settlement contracts immutably anchor the updated state roots and balance proofs, completing the loop between fast off-chain execution and on-chain verifiability.

zkSpot demonstrates how Chainless Apps eliminate traditional trade-offs. It offers CEX-level UX without compromising Web3-grade guarantees — and without launching a dedicated rollup or validator set. Developers simply plug into the modular Agglayer infrastructure. The same architectural principles can be applied to unlock many other types of applications.

\subsection{Other Chainless Designs}

While zkSpot highlights one high-trust model (TEE + zkVM), the Chainless framework unlocks a broad range of applications using different execution and trust configurations — tailored for their domain-specific needs.

\textbf{Verifiable web3 games}:  
A turn-based multiplayer game runs its logic off-chain in native runtime. Each player’s actions (e.g., move, attack) are recorded into state transitions and batched as execution traces. A zkVM verifies that all game logic was followed correctly — enforcing fairness without requiring all logic to run on-chain. Agglayer interop allows NFT rewards or state syncs to flow across chains. This design enables games to feel real-time and interactive while remaining cryptographically verifiable.

\textbf{Private institutional finance systems}:  
A bank or regulated institution can deploy a private Chainless App to manage internal processes like payments, compliance checks, or trade settlement. The Execution Layer runs business logic off-chain in a secure enclave (TEE), while trust is enforced through internal attestations, audit logs, or selectively applied zk proofs. In this setting, the Trust Layer may include formal compliance attestation rather than zkVM verification, or use optimistic proofs with legal recourse. If public visibility is needed (e.g., for regulatory settlement), finalized states or aggregate proofs can be posted on-chain via Agglayer. This architecture unlocks permissioned systems that retain auditability, verifiability, and interoperability—without requiring a full blockchain stack.

\textbf{Verifiable Web2 integration via zkTLS}:  
A Chainless App can securely ingest data from HTTPS APIs (e.g., price feeds, off-chain identities, or attestations) using zkTLS. In this setup, the Execution Layer establishes a TLS connection with a Web2 service and captures the server’s signed response. The Trust Layer then generates a zero-knowledge proof that this response was delivered over a valid TLS session, including verification of the TLS certificate chain and server authenticity, without revealing the full session content. This proof is generated using a zkVM-based TLS prover, and can be posted on-chain or passed across chains via Agglayer. This unlocks verifiable oracles and Web2–Web3 bridges that operate without trusted intermediaries, enabling trustless consumption of HTTPS data in smart contracts.

These examples highlight the expressive power of Chainless Apps. Developers can pick the right mix of performance, verification, and interoperability tools — with Agglayer enabling seamless cross-chain interaction regardless of the internal trust model.
\section{Conclusion}

Chainless Apps present a new modular paradigm for building blockchain applications that combine the performance of Web2 systems with the verifiability and decentralization guarantees of Web3. By decoupling execution, trust verification, cross-chain interoperability, and settlement into distinct layers, Chainless Apps allow developers to optimize for scalability, flexibility, and security independently at each stage.

Central to this design is app-specific sequencing, where each application maintains independent control over transaction ordering and state evolution off-chain. This enables domain-optimized performance while retaining the ability to later verify and settle results trustlessly on-chain.

Through verifiable compute techniques such as zkVMs, validator committees, or trusted hardware, applications can run high-throughput operations off-chain while maintaining cryptographic or decentralized trust guarantees. Agglayer further strengthens this model by providing a standardized interoperability and settlement infrastructure — enabling seamless asset and message bridging, liquidity unification across chains, and real-time cross-chain communication, all anchored by pessimistic proofs for safety.

This architecture allows developers to select trust models that fit their specific use cases, ranging from high-assurance zk verification to lightweight committee or legal enforcement models, without fragmenting the broader security and liquidity ecosystem. It also ensures that even applications with diverse internal assumptions can interoperate safely through Agglayer’s shared settlement framework.

The zkSpot example demonstrates the viability of Chainless Apps today: achieving CEX-like UX while preserving Web3-grade guarantees — without the need to deploy new blockchains, manage validator sets, or fragment liquidity. Moving forward, the Chainless App framework has broad applicability beyond trading into gaming, social platforms, data availability, and beyond — offering a new foundation for scalable, trust-minimized, and interoperable blockchain applications.

In a modular and multichain future, Chainless Apps, powered by app-specific sequencing and cross-chain interoperability, represent a fundamental step toward realizing the vision of a seamless, verifiable, and user-centric Web3 ecosystem. The vision aligns with modular decentralization pathways\cite{Buterin2021Endgame}.
\section{Acknowledgements}

The authors would like to thank the broader Polygon Labs team for insightful discussions on modular blockchain design, verifiable computation, and cross-chain interoperability that helped shape the ideas presented in this paper. Special thanks to the developers and researchers behind projects such as Polygon Zero(Plonky3), Polygon Miden, RISC Zero, EigenLayer, Succinct, and Flashbots, whose pioneering work on zkVMs, decentralized validation, and MEV minimization provided critical technical foundations. 

We would like to specifically acknowledge David Silverman for contributing the ``Just Trust Me Bro'' (JTMB) terminology, which is Operator Trust and for valuable discussions around the early concepts of Chainless Apps. We also thank Pablo Vigara and Saúl García for their work on implementing the zkHyperliquid(zkSpot) proof-of-concept demo, which helped demonstrate the practical viability of the Chainless App architecture. We also extend our appreciation to Rongxin Zhang from Whitepaper Reading Club, Ellie Davidson from Espresso Systems, Katie Mckeon and Stasia Carson from Sindri team, Soubhik Deb and Ishaan 0x from EigenLayer team, last but not least Abhishek Agarwal and Brendan Farmer from Polygon Labs for the extensive review. 

Finally, we acknowledge the broader modular blockchain and Layer 2 research community, including the authors of foundational works on zkSNARKs, zkSTARKs, rollup architectures, and application-specific sequencing, for advancing the field and inspiring the Chainless App framework. We appreciate the open-source and academic communities whose contributions have made this work possible.

\end{multicols}

\end{document}